# Comb-based photonic neural population for parallel and nonlinear processing

BOWEN MA, JUNFENG ZHANG, AND WEIWEN ZOU*

*State Key Laboratory of Advanced Optical Communication Systems and Networks, Intelligent Microwave Lightwave Integration Innovation Center (imLic), Department of Electronic Engineering, Shanghai Jiao Tong University, 800 Dongchuan Road, Shanghai 200240, China*
*Corresponding author: wzou@sjtu.edu.cn*

**It is believed that neural information representation and processing relies on the neural population instead of a single neuron. In neuromorphic photonics, photonic neurons in the form of nonlinear responses have been extensively studied in single devices and temporal nodes. However, to construct a photonic neural population (PNP), the process of scaling up and massive interconnections remain challenging considering the physical complexity and response latency. Here, we propose a comb-based PNP interconnected by carrier coupling with superior scalability. Two unique properties of neural population are theoretically and experimentally demonstrated in the comb-based PNP, including nonlinear response curves and population activities coding. A classification task of three input patterns with dual radio-frequency (RF) tones is successfully implemented in a real-time manner, which manifests the comb-based PNP can make effective use of the ultra-broad bandwidth of photonics for parallel and nonlinear processing. © 2021 Optical Society of America**

## 1. INTRODUCTION

Increasing evidence in neuroscience supports that the neural population plays a critical role in robust information representation, communication, and processing [1]. As an ensemble of neurons and synaptic interconnections, it provides a deeper understanding of the dependence between global functions and elementary details. The neural population features the nonlinear response curve of individual neurons (i.e., bell-shape tuning curves or Gaussian receptive fields) and activities coding at population level [2]. The former reveals that a neuron is interested in a part of the input range and responds to it in a distributed and nonlinear way [3]. The latter indicates population activities constructed by these responses can be regarded as state variables corresponding to input patterns [4]. In recent years, the advances in electronic neuromorphic computing indicate the way for these properties to support population-based information processing [5]. Required nonlinear responses are built up by the weighted sum of a set of superparamagnetic basic functions that agree well with the bell-shape tuning curves [6]. The input vowel patterns can be mapped onto the population activities of four coupled spin-torque nano-oscillators for efficient recognition [7]. Generally, the scalability and interconnection of neurons are pre-requisites for accessing a population-based information processing.

However, it remains challenging to achieve a photonic neural population (PNP) equipped with nonlinear response curves and efficient population coding. In neuromorphic photonics, the unique advantages of broad bandwidth and low power consumption have dominated high-speed and massively-parallel linear computations (i.e., synaptic weighting) [8]. Numerous studies also focus on the implementations of nonlinear computations (i.e., the neuron activation) based on single devices or temporal nodes, as illustrated in Fig. 1(a) [9]. Specifically, we attribute the challenge of establishing a PNP to the physical complexity, response latency, and reconfigurability. First, the complicated implementation of a single-device-based photonic neuron limits the scalability and interconnection, for example schemes with a laser [10-15], a modulator [16], and a resonator [17,18]. Second, the scaling up of a temporal photonic neuron introduces unwanted response latency, which may counteract the advantages of a PNP for high-speed processing [19,20]. Third, the reconfigurability of a PNP is critical for broad processing tasks, which is challenging for the photonic neurons based on the embedded nonlinear material with fixed topology [21].

Here, we propose a comb-based PNP to enable a population-based scheme for parallel and nonlinear processing. The comparison of the comb-based PNP and its biological counterpart is conceptually illustrated in Figs. 1(b-c). It is observed that an injected optical frequency comb (OFC) features inverse bell-shape nonlinear responses to input frequency components and that they are reconfigurable through the OFC parameters. Therefore, these distributed responses support a high-dimensional representation of complex input patterns through the comb power distribution (i.e., population activities). It is experimentally indicated that the interconnection between the comb neurons is naturally provided by the carrier coupling. It is found that the lateral inhibition enhances the response nonlinearity and leads to a unique population activity. We also implement a classification task of three input patterns with dual radio-frequency (RF) tones. The distinct population activities of the comb-based PNP can be efficiently read out for classification. For the comb-based PNP, the inherent broad bandwidth of photonics ensures superior scalability and the ability of nonlinear processing of RF input patterns in real time comparted to other neuromorphic and conventional photonics-assisted paradigms [22].

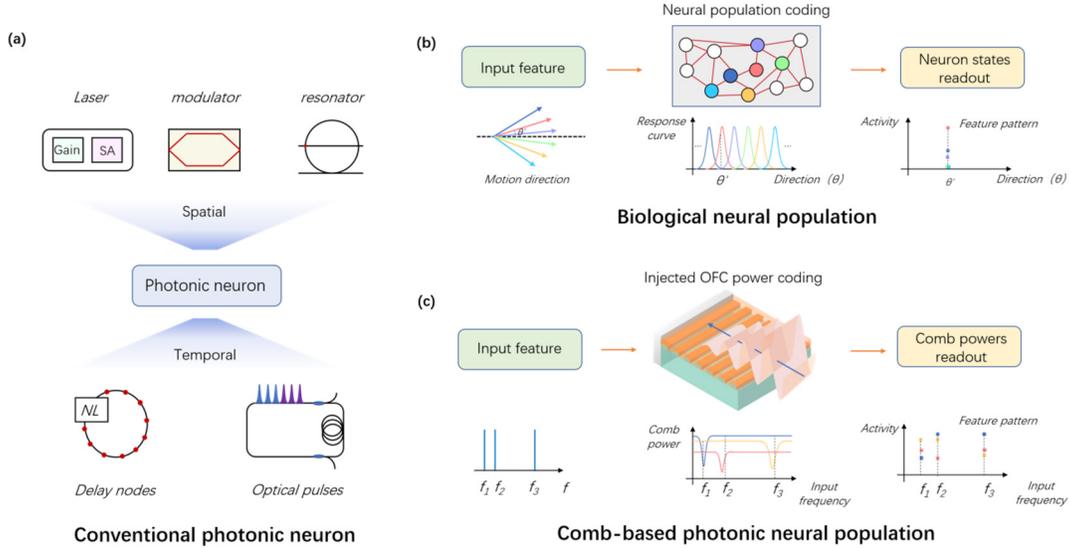

**Fig. 1.** Conceptual illustrations of conventional photonic neurons, biological neural population, and the proposed comb-based photonic neural population (PNP). (a) Existed photonic neurons scale up in spatial domain or temporal domain at the cost of physical complexity or response latency. The former is based on a single structure or devices, such as lasers [10-15], modulators [16], and resonators [17,18]. The latter relies on the temporal nodes, such as in reservoir computing [19] and optical pulses in ising machine [20]. (b) Schematic of the biological neural population with an input of motion direction. The neurons respond to different parts of input range in a distributed manner (known as the bell-shape tuning curves). Complex input patterns can be represented by neural population coding and classified by an efficient readout process. (c) The proposed comb-based PNP. An injected optical frequency comb (OFC) can nonlinearly respond to input patterns in frequency domain. Photonic tuning curves and comb power coding are exploited for processing complex input patterns. The comb powers can be obtained by a one-time readout, which ensures a low response latency.

## 2. Methods

### A. Theory

The comb-based PNP includes population generation, parallel processing, and population activity reading out, corresponding to the implementation of a dual-OFC generation, the injected OFC to interact with the input, and a beat process for the dual-OFC, respectively. The scalability and tunability of the OFC [23] help with the flexible control of the PNP properties at both a neuron-level (e.g., initial comb power) and a population-level (e.g., comb spacing). The monitoring of the population activities benefits from the introduction of another coherent OFC [24] and a beat process. The process is described below. First, the electro-optical dual-OFC is prepared at a comb spacing of $f_1$ and $f_2$ with a slight frequency difference of $\Delta f$ and a comb number of $N_c$. One of the OFCs undergoes a frequency shift operation of $f_0$ to make use of all available combs as state variables in the later beat process. Second, the frequency-shifted OFC is injected to a distributed-feedback laser diode (DFB-LD) with a central wavelength at $\lambda_0$ and the detuning frequency is noted as $f_{detu}$. The input frequency at $f_{in}$ is modulated on the pump current of DFB-LD. Lastly, the output of DFB-LD is coupled with the other OFC for a beat process in a photodetector (PD). The beat note powers at $f_0 \pm h \cdot \Delta f$ ($h$ is a natural number) are monitored as signatures of the intensity of injected combs, whose distribution has been changed by input frequency.

The injection dynamics of an OFC to semiconductor lasers (SL) dependent on various injection ratios and detuning frequencies was theoretically and experimentally investigated [25,26]. We concentrate on the effect of input frequency on the comb power under different configurations of the OFC and DFB-LD. A theoretical model with a modified term of pump current is used for the numerical simulation (see Appendix A). In the optical cavity, the injected OFC and input frequency act as two sources of carrier modulation to drive the resonance of the output optical field. As a result, they interact with each other through carrier coupling, leading to an input-frequency-dependent variation of the comb power. In numerical investigations, we apply a scanning input frequency and record the comb power to study the shape and width of the response curve, which are related to the DFB-LD bias current, the input RF power, and the initial comb power (see Appendix B). Furthermore, we simulate the responses of 15 comb powers to 4 complex input patterns with 15 frequency tones to demonstrate the representation and classification ability of the comb-based PNP (see Appendix C). To investigate the comb power distributions of similar input frequencies, the actual input frequencies consist of accurate frequencies and a random variation value.

### B. Experimental demonstration

The experimental setup of the comb-based PNP is presented in Fig. 2. A schematic of the PNP is shown on the upper side. The node colors of different layers are in accordance with the background color of experimental stages on the lower side. The PNP is generated by an electro-optical dual-OFC scheme, where the output of a narrow-linewidth continuous-wave laser (CW, TLG-200, Alnair Labs) is split into two branches to be respectively modulated through phase modulators (PM, PM-0K5-10-PFA-PFA, EOspace). Then the central wavelength of one OFC is shifted for 100 MHz by an acousto-optical modulator (AOM, SGTF100-1550-1, Smart Sci & Tech). Using two phase-locked RF channels of a signal generator (N5183B, Keysight), the modulation frequencies noted as $f_1$ and $f_2$ correspond to the dual-comb spacings and are set as the upper limit of the response curve width of a single comb neuron. Note that more combs can be obtained to enlarge the PNP through a larger output power of RF signals by low-noise amplifiers (LNA, LNA-3035 & LNA-3020, RF Bay) or more complex architectures with cascaded modulators [27]. Consequently, the comb spacing and number can be flexibly tuned for different response coverages of the comb-based PNP. As discussed later, the flatness of the OFC is unnecessary since the initial comb power can influence the response curve width (e.g., biological neurons with diverse areas of receptive fields).

The frequency-shifted OFC is injected into a DFB-LD (T-20-550-B-07#G, Henan Shijia) with no optical isolator at the output port through an optical circulator and a polarization controller (PC). A DFB-LD with a modulation bandwidth of 20-GHz is provided for input RF signals. As a result, the carrier coupling provides the inter-layer connection (i.e., layer i and layer ii) and the intra-layer connection (i.e., inside layer i). It is known that both the optical injection and direct modulation can introduce a red shift of the central wavelength of the DFB-LD, which is dependent on the injection power or modulation depth [28]. Hence, a relatively low input power (SMA-100B, Rohde & Schwarz) can reduce the effect of the unwanted detuning wavelength. Another advantage is the elimination of modulation harmonics, which may disturb the distributed response of the comb neuron. The output of the DFB-LD is divided into two parts for monitoring by an optical spectrum analyzer (OSA, AQ6370C, Yokogawa) and coupling with the other OFC for heterodyne interference, respectively. The electric signal of a 500-MHz PD (FS-PD-B-2030, fsphotonics) laid after the heterodyne interference is monitored by an electrical spectrum analyzer (ESA, FSW, Rohde & Schwarz) centered at 100 MHz (i.e., the shifted frequency by AOM) with a 500-Hz resolution bandwidth for the readout of the comb power variation through the beat notes at $100 \pm h \cdot \Delta f$ MHz (i.e., the connection between layers i and iii).

Given that the CW laser and DFB-LD operate independently with noise and temperature drift, it is critical to keep their detuning frequency (i.e., $f_{detu}$) stable versus time. A pre-calibration method is implemented to introduce an increment of the comb power at the lowest sideband as an indication of $f_{detu}$ (see Appendix D). This method facilitates the monitoring and controlling of the $f_{detu}$ for a reliable response range of the comb-based PNP.

## 3. Results

### A. Photonic tuning curve of comb neurons

The responses of a comb neuron to a single input frequency are given in Fig. 3. The CW laser is centered at the red-side of DFB-LD with a wavelength difference of 0.1 nm and an optical output of 15 dBm. We apply a 25-dBm RF power for dual-OFC generation with 6 beat notes as the comb neuron signatures. The $f_1$ and $f_2$ are 3 GHz and 2.989 GHz, respectively. As indicated above, the pre-calibration method is implemented with a 31.9-mA bias current of DFB-LD. The power of input frequency is also as low as -7 dBm.

Figures 3(a-d) present the schematics of response cases corresponding to various input frequency bands. We focus on the central comb neuron and the adjacent two comb neurons, denoted as comb 0, comb -1, and comb 1, respectively. Besides the no-input case in Fig. 3(a), input frequencies fall into separated response ranges of different comb neurons (i.e., the pink background), as shown respectively in Figs. 3(b-d). Figures 3(e-h) and 3(i-l) correspond to the optical spectrums and electrical spectrums in each case, where the comb neuron responses are highlighted by a dotted box in red. In Fig. 3(e), there are three distinguished peaks of the injected OFC. The asymmetric power distribution of beat notes

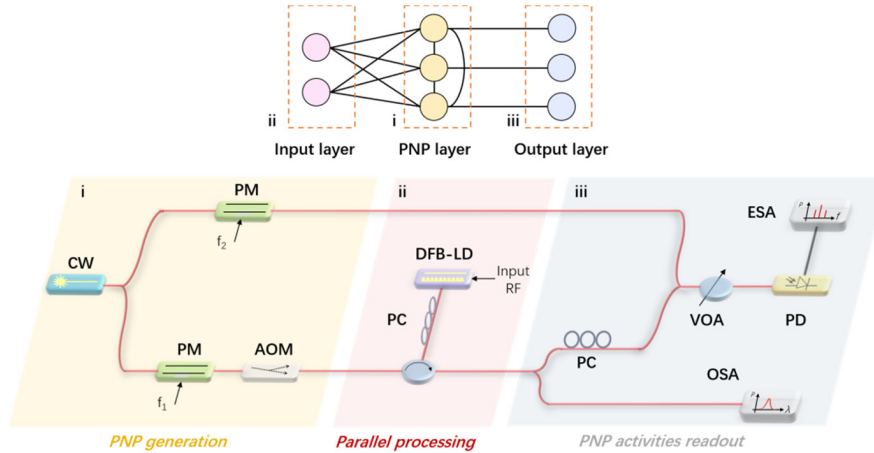

**Fig. 2.** Experimental setup and corresponding schematic of the comb-based PNP. On the upper side, the input layer connects to the PNP layer by carrier coupling. The gain competition inside the PNP resembles the lateral inhibition connection in the biological counterpart [29]. The output layer represents the beat notes for the PNP activities monitoring. On the lower side, an electro-optical dual-OFC with slightly different comb spacings is generated by parallel PMs. The frequency-shifted OFC is optically-injected into a DFB-LD and the PC is used for injection efficiency optimization. The output of the DFB-LD is monitored by an OSA and then detected in a low-speed PD with the other OFC. CW: continuous-wave laser, PM: phase modulator, AOM: acousto-optical modulator, PC: polarization controller, DFB-LD: distributed-feedback laser diode, OSA: optical spectrum analyzer, VOA: variable optical attenuator, PD: photodetector, ESA: electrical spectrum analyzer.

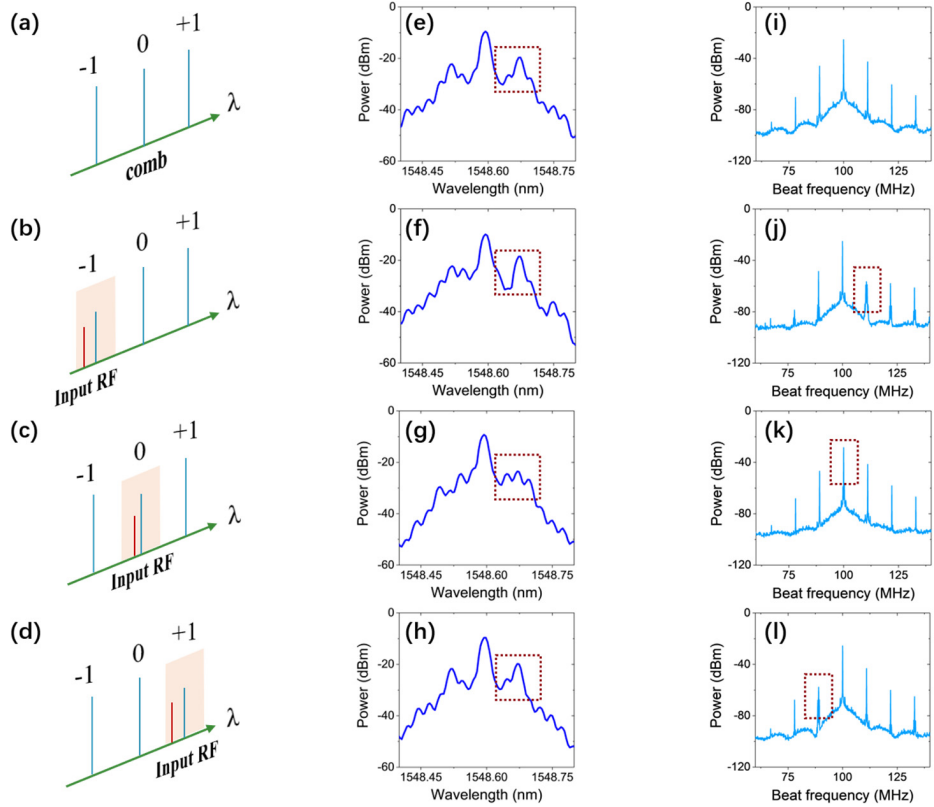

**Fig. 3.** The schematics and corresponding experimental results of the comb neuron response for a single input frequency. (a-d) The illustrative cases of no-input, 6.6-GHz-input, 9.6-GHz-input, 12.4-GHz-input, respectively. The input frequency falls into response ranges of respective comb neurons and leads to a variation of the comb power. (e-h) Optical spectrums of each input case. The dotted box in red highlights the apparent decrease of corresponding comb power. (i-l) Electrical spectrums of each input case. The comb powers are monitored through the beat note powers. The center frequency and the spacing of the beat notes correspond to the frequency shift and frequency difference of the OFCs, respectively.

is a result of the in-cavity optical gain as shown in Fig. 3(i). As a consequence of the pre-calibration method, the response ranges are around the input frequencies of multiple times of $f_1$ (see Appendix D). The responses to an input frequency at 6.6 GHz, 9.6 GHz, and 12.4 GHz are illustrated in Figs. 3(f-h) and 3(j-l), respectively. An evident power decrease in the corresponding comb neuron is sequentially observed in both optical and electrical spectrums. It is demonstrated that the interaction between in-cavity frequency components leads to a power variation of the comb neuron in response to an input frequency of different bands.

We scan the input frequency to investigate the response curve of comb neurons. In Fig. 4, the beat note powers corresponding to different input frequencies are given. First, we find that the distributed response of respective comb neurons resembles an inverse bell-shape tuning curve. It demonstrates the neural-like response behavior of the comb neuron. Second, the width and amplitude (i.e., the maximum variation of comb power) of the response curve are dependent on the initial comb power. A relatively flat response curve of comb 0 shows a larger response width and a lower response amplitude compared to those of comb 1 and comb -1. The comparable initial powers of comb 1 and comb -1 lead to their similar response widths and amplitudes. Third, gain competition effect is observed that the comb -1 and comb 1 increase when comb 0 decreases. This is a reminiscence of the well-known lateral inhibition mechanism between biological neurons [29] which can highlight the feature contrast.

Time stability of the comb neuron response is studied by comparing the beat note powers under identical input after a period of time (see Fig. 8(c)). The stable response manifests the feasibility of the pre-calibration method for eliminating the effect of drifting $f_{detu}$ and ensures the population-based information processing.

Further investigation on the shapes of response curve is presented and discussed in Appendix B. The numerical results of the inverse bell-shape tuning curve are shown in Fig. 6(a), which agree well with the experimental results in Fig. 4. Besides, other shapes of response curve with both positive and negative peaks are also theoretically and experimentally observed (see Figs. 6(b-e)). The comb power starts to rise as the input frequency scans, then falls below the initial level, and gradually recovers to the initial level. It indicates the comb neurons can provide both excitatory and inhibitory responses to an input frequency. Moreover, it is found that the response width and amplitude are also dependent on the input RF power as presented in Fig. 6(f).

### B. Comb power coding for input pattern classification

A classification task of three input patterns with dual RF tones is implemented with the comb-based PNP. An arbitrary waveform generator (AWG, M8195A, Keysight) outputs a single RF tone with an amplitude of 600 mV, which is subsequently filtered by a band-pass filter (BPF, BPF6-10Gs-2976, RF bay) to remove the spurs.

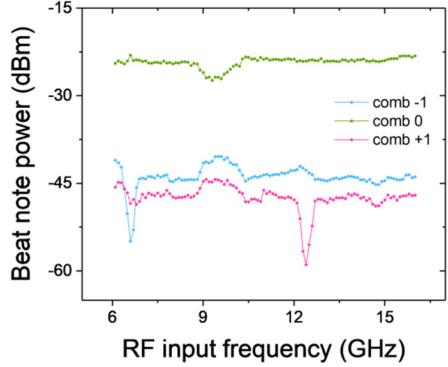

**Fig. 4.** The beat note powers dependent on input frequencies. The response range locates around the multiple times of the comb spacing. The inverse bell-shape response curve of each comb neuron resembles the tuning curve of biological neurons. The response width and amplitude are related to the initial comb power.

Another 4-dBm RF tone is produced by the RF signal generator and combined with the output of BPF as the input of DFB-LD.

We prepare the inputs through a combination of random tones around the three frequency bands of 6 GHz, 9 GHz, and 12 GHz corresponding to the response ranges of the comb neurons. The experimental results of the relative beat note powers are shown in Fig. 5. The faded circles represent the PNP activities for each input. The stars indicate the average activities for three input patterns that are in red, green, and purple. Additionally, the projections of the average activities to the planes are marked. It is found that every input pattern occupies adjacent positions in this high-dimensional state space created by the comb powers. It manifests that the comb-based PNP has approximate responses to similar inputs. For example, when band 1 and band 2 are involved in the input, beat power 1, beat power 2, and beat power 3 decreases, rises, and remains invariant, respectively. Consequently, the comb powers function as a set of state variables mapping onto the input pattern and the PNP activities form three distinct clusters, which can be read out for classification through simple linear regression in a nearly real-time manner. Note that the nonlinear and distributed response of the comb neuron is critical to the classification ability of the comb-based PNP.

As the number of comb neurons and input frequencies increases, the response of the PNP is non-trivial considering the intracavity interaction and effects like gain competition, which may lead to more complex and unique PNP activities. In the numerical investigations, the PNP activities of 15 comb neurons for 4 input patterns with 15 RF tones are compared (see Appendix C). The results show that a comb power increases (i.e., an excitatory response) if there is an RF tone around itself, which means the comb-based PNP can process the input pattern in a parallel and distributed style. Therefore, the input patterns are successfully classified according to the PNP activities.

## 4. Conclusion

We have demonstrated a comb-based PNP to achieve a population-based paradigm for parallel and nonlinear processing. The implementations of population generation, parallel processing, and population activity reading out are introduced. Furthermore, we have presented the distributed response of comb neurons as photonic tuning curves, and comb power coding for input pattern classification. Three features are highlighted for the comb-based PNP. First, it keeps a low physical complexity toward a large population with the PNP-tunability and activities-readability. The promising OFC technology ensures a chance of the comb-based PNP to scale up with abundant bandwidth resources [23]. The PNP can also be efficiently tuned and read out instead of accessing to individual neuron one by one. Second, the carrier-coupling naturally creates the complex interconnection with compactness. A similarity between gain competition and lateral inhibition is interesting for potential higher-level population-based processing [30]. Third, it enables a neuromorphic RF processing in real time, which indicates a broader prospect of neuromorphic photonics besides the popular paradigms such as optical neural network (ONN) [31]. The inherently parallel and distributed style of photonics and neuromorphic processing may provide inspirations for shared processing principle. To take a step further, we consider the comb-based PNP in the context of the conventional photonics-assisted RF signals characterization [22]. Generally, a linear mapping is constructed between the RF domain and spatial or temporal domain [32,33]. It introduces increased physical complexity or extra response latency (i.e., a scanning period). In applications such as spectrum monitoring and ultra-fast frequency-hopping communication, the response latency of RF signals characterization is critical for the performance.

The comb-based PNP provide a real-time and compact approach to RF signals characterization. Note that the nonlinear processing greatly improves the characterization efficiency. For example, when dealing with broadband and sparse RF signals, the conventional schemes are less efficient as most of the time resources and channels are wasted without obtaining valuable information. However, the comb-based PNP can directly sense the dynamical change of the RF signals with just a series of comb powers. In future studies, the PNP generation can be improved to generate more comb neuros and equipped with a programmable power distribution through a wave-shaper. The pre-calibration method can be implemented with a feedback control to efficiently guarantee the stability. It is also interesting to study the PNP interconnection dependent on cavity properties for a better understanding of the network topology. At last, recent breakthroughs of substantially-efficient on-chip OFC [34,35] indicate the potential monolithic integration of the comb-based PNP.

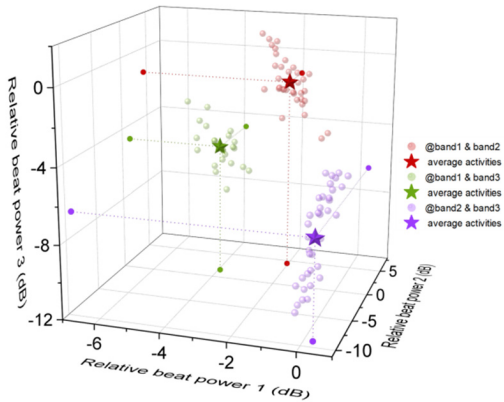

**Fig. 5.** The comb-based PNP activities of three input patterns with dual radio-frequency (RF) tones. The tones are randomly selected from three frequency bands. The PNP shows similar activities in response to the input pattern with approximate frequencies. Classification results can be consequently read out.

**APPENDIX A: THEORETICAL MODEL**

The dynamics of a SL when it is injected by an OFC and simultaneously driven by an input RF signal are described. The dynamics characterization is based on the amplitude $E(t)$, phase $\varphi(t)$ of the electrical field of injected laser, and the carrier density $N(t)$. Following the model in recent studies [25,26], we modify the pump current term with a modulation expression of $R_p(t) = R_b + m \cdot (R_b - R_{th}) \cdot \sin(2\pi f_{in} t)$, where $R_p$, $R_b$, $R_{th}$, $f_{in}$ represent pump rate, pump rate bias, pump rate at threshold, and modulation frequency, respectively. The power of input RF signal is adjusted by $m$. In simulations, SL parameters are chosen as the following: carrier lifetime $\tau_s = 2\times10^{-9}$s, differential gain $G_N = 7.9\times10^{-13}$m$^3$s$^{-1}$, threshold carrier density $N_{th} = 2.91924\times10^{24}$m$^{-3}$, photon lifetime $\tau_p = 2\times10^{-12}$s, linewidth enhancement factor $\alpha = 5$, and pump rate at threshold $R_{th} = 1.8\times10^{33}$m$^{-3}$s$^{-1}$. The main factors of comb spacing $f_1$, comb number $N_c$, detuning frequency $f_{detu}$ between the nearest comb and SL, and relative injection amplitude $E_{inj}$ of each comb, are variable. Consequently, we observe the optical spectrum of the complex electrical field of injected laser by discrete Fournier transform.

**APPENDIX B: CHARACTERIZATION OF PHOTONIC TUNING CURVE**

To characterize the response of the comb power to a single input frequency (i.e., photonic tuning curve), we scan the $f_{in}$ with a frequency step of 50 MHz and record corresponding comb peak powers. In Fig. 6, two typical shapes of the photonic tuning curve are observed, including the inverse bell-shape and the dual-peak shape. The numerical results are presented in the Figs. 6(a-c). We set up 3 combs with a 2-GHz spacing. As shown in Fig. 6(a), applying $R_b = 1.2R_{th}$, $m = 0.05$, $E_{inj} = 0.3$, and $f_{detu} = 100$ MHz, the response curve has an inverse bell-shape, which agrees well with that in Fig. 4. In Fig. 6(b), the response curve features with a positive peak and a negative peak, for $R_b = 4R_{th}$, $m = 0.1$, $E_{inj} = 0.1$, and $f_{detu} = 2.01$ GHz. The response curve of all three combs is given in Fig. 6(c) under the same condition as those in Fig. 6(b) except for $m = 0.3$. We find that there is a higher positive peak and a larger response width than those in Fig. 6(b). The experimental characterization of photonic tuning curve is shown in Figs. 6(d-f). The dual-peak shape is also observed in Fig. 6(d) with a 1-GHz comb spacing. Figure 6(e) gives another detailed dual-peak response curve with a large response width of ~5 GHz. It is found that the slope of the response curve gradually increases. We also investigate the effect of the input power as shown in Fig. 6(f). The results indicate the positive peak increases as the input power, which is consistent with the comparison among Figs. 6(b-c). A large input power of ~15 dBm leads to an obvious red shift of the cavity. In general, the competition for carrier between the comb and input frequency is dependent on their relative intensity, detuning frequency, and pump rate bias, leading to a distributed and nonlinear photonic tuning curve.

**APPENDIX C: CONFIGURATION OF COMPLEX PATTERNS CLASSIFICATION**

To numerically investigate the response to complex input patterns, we consider a classification task of 4 input patterns of 15 input tones, and record the optical spectrum of 15 combs separated by 150 MHz, as shown in Fig. 7(a). Applying $f_{detu} = 2.01$ GHz, the frequency range from 1.855 GHz to 4.105 GHz is divided into 15 input frequency bands corresponding to 15 combs. We regard the input frequencies falling into the same band as an identical feature for the overall input pattern. Each input frequency is defined as a combination of an ideal value and a random value with a ratio of 0.01. The binary formats of the 4 input patterns are selected as: [0, 1, 0, 1, 0, 1, 1, 0, 1, 1, 1, 1, 1, 0, 1] of pattern 1, [0, 1, 1, 1, 0, 1, 1, 1, 0, 1, 0, 1, 0, 1, 1] of pattern 2, [1, 1, 1, 1, 0, 1, 1, 0, 0, 1, 0, 0, 1, 1, 1] of pattern 3, and [1, 1, 0, 1, 0, 1, 1, 0, 1, 1, 0, 1, 0, 1, 1] of pattern 4, where the binary 1 indicates that there is a tone inside the corresponding input frequency band. The numerical optical spectrums are shown in Figs. 7(b-e). A dotted line helps with a separation of the comb power level into two categories. The sequence numbers of the lower comb powers are marked. We find that a comb power increases if there is an input tone within the frequency band. There is a good agreement between the comb power distribution and the complex input pattern. It is indicated that the comb-based PNP can classify RF signals in a distributed and parallel manner.

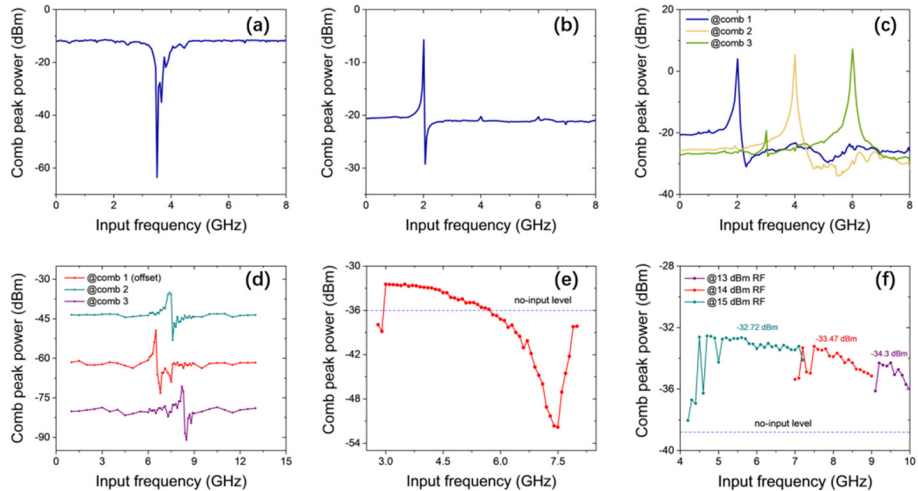

**Fig. 6.** The characterization of shapes of the photonic tuning curve. (a-c) Numerical results of the inverse bell-shape, dual-peak shape, distributed dual-peak shape, respectively. (d-f) Experimental results of the distributed dual-peak shape, dual-peak shape, and input-power dependence.

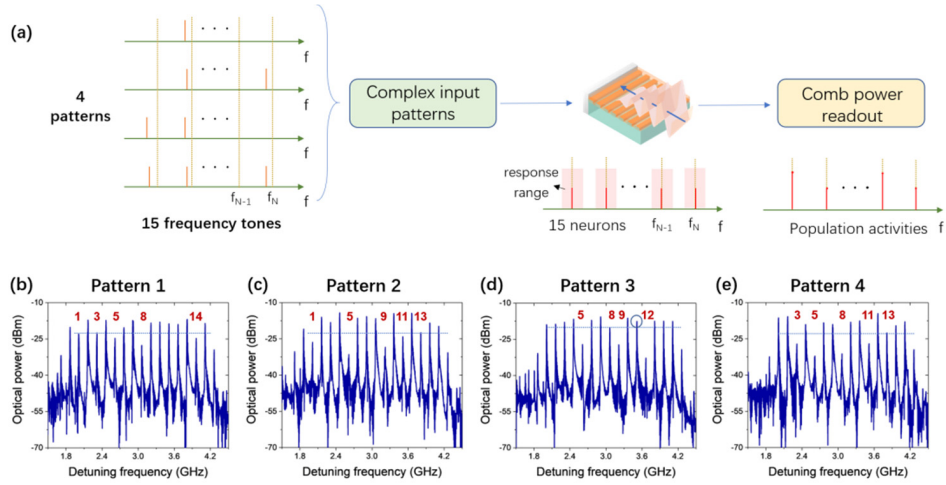

**Fig. 7.** Numerical results of classification of the complex input patterns. (a) Schematic of the classification task of 4 input patterns with 15 tones. (b-e) Optical spectrums corresponding to the 4 input patterns, respectively. The sequence numbers of the lower comb powers are indicated.

## APPENDIX D: PRE-CALIBRATION METHOD

We implement a pre-calibration experimental method to overcome the drifting of $f_{detu}$. First, we adjust the center wavelength of the DFB-LD to approach the nearest comb by tuning the bias current. The power of the nearest comb (i.e., state comb) tends to increase as $f_{detu}$ decreases. Second, we select a bias current value corresponding to a stable and high state comb power. As shown in Figs. 8(a-b), we record the power value as a signature of $f_{detu}$. Third, we monitor the state comb power and finely adjust the bias current to ensure a stable $f_{detu}$. As a result, the response range of a comb locates around the multiple times of the comb spacing, which provides a design reference to the comb-based PNP for different processing tasks. Figure 8(c) presents the response stability versus time corresponding to Fig. 4. With the pre-calibration method, the three combs reliably response to the same input frequency.

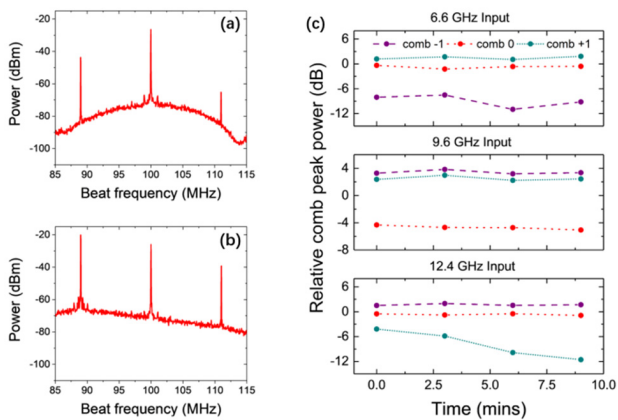

**Fig. 8.** The experimental results of pre-calibration method. (a, b) The state comb power increases as the detuning frequency ($f_{detu}$) decreases. Hence, the comb power can be used as a sign of $f_{detu}$. (c) Time stability of the comb response corresponding to the Fig. 4.

**Funding.** National Key Research and Development Program of China (Program No. 2019YFB2203700); National Natural Science Foundation of China (Grant No. 61822508).

**Disclosures**. The authors declare no conflicts of interest.

**Data availability**. Data underlying the results presented in this paper are not publicly available at this time but may be obtained from the authors upon reasonable request.

## References

1. W. Ma, J. M Beck, P. E Latham, and A. Pouget, "Bayesian inference with probabilistic population codes," Nat. Neurosci. **9,** 1432 (2006).
2. T. Womelsdorf, K. Anton-Erxleben, F. Pieper, and S. Treue, "Dynamic shifts of visual receptive fields in cortical area MT by spatial attention," Nat. Neurosci. **9,** 1156 (2006).
3. A. Pouget, P. Dayan, and R. Zemel, "Information processing with population codes," Nat. Rev. Neurosci. **1,** 125 (2000).
4. K. Roth, S. Shao, and J. Gjorgjieva, "Efficient population coding depends on stimulus convergence and source of noise," PLoS Comput. Biol. **17,** e1008897 (2021).
5. J. Grollier, D. Querlioz, K. Y. Camsari, K. Everschor-Sitte, S. Fukami, and M. D. Stiles, "Neuromorphic spintronics," Nat. Electron. **3,** 360 (2020).
6. A. Mizrahi, T. Hirtzlin, A. Fukushima, H. Kubota, S. Yuasa, J. Grollier, and D. Querlioz, "Neural-like computing with populations of superparamagnetic basis functions," Nat. Commun. **9,** 1533 (2018).
7. M. Romera, P. Talatchian, S. Tsunegi, F. A. Araujo, V. Cros, P. Bortolotti, J. Trastoy, K. Yakushiji, A. Fukushima, H. Kubota, S. Yuasa, M. Ernoult, D. Vodenicarevic, T. Hirtzlin, N. Locatelli, D. Querlioz, and J. Grollier, "Vowel recognition with four coupled spin-torque nano-oscillators," Nature **563,** 230 (2018).
8. G. Wetzstein, A. Ozcan, S. Gigan, S. Fan, D. Englund, M. Soljačić, C. Denz, D. A. B. Miller, and D. Psaltis, "Inference in artificial intelligence with deep optics and photonics," Nature **588,** 39 (2020).
9. B. J. Shastri, A. N. Tait, T. F. de Lima, W. H. P. Pernice, H. Bhaskaran, C. D. Wright, and P. R. Prucnal, "Photonics for artificial intelligence and neuromorphic computing," Nat. Photonics **15,** 102 (2021).
10. B. Ma and W. Zou, "Demonstration of a distributed feedback laser diode working as a graded-potential-signaling photonic neuron and its

application to neuromorphic information processing," Sci. China Inf. Sci. **63,** 160408 (2020).
11. M. A. Nahmias, B. J. Shastri, A. N. Tait, and P. R. Prucnal, "A leaky integrate-and-fire laser neuron for ultrafast cognitive computing," IEEE J. Sel. Top. Quant. **19,** 1800212 (2013).
12. S. Xiang, Z. Ren, Z. Song, Y. Zhang, X. Guo, G. Han, and Y. Hao, "Computing primitive of fully VCSEL-based all-optical spiking neural network for supervised learning and pattern classification," IEEE T. Neur. Net. Lear. **32,** 2494 (2020).
13. G. Sarantoglou, M. Skontranis, and C. Mesaritakis, "All optical integrate and fire neuromorphic node based on single section quantum dot laser," IEEE J. Sel. Top. Quant. **26,** 1900310 (2020).
14. V. A. Pammi, K. Alfaro-Bittner, M. G. Clerc, and S. Barbay, "Photonic computing with single and coupled spiking micropillar lasers," IEEE J. Sel. Top. Quant. **26,** 1500307 (2020).
15. J. Robertson, E. Wade, Y. Kopp, J. Bueno, and A. Hurtado, "Toward neuromorphic photonic networks of ultrafast spiking laser neurons," IEEE J. Sel. Top. Quant. **26,** 7700715 (2020).
16. A. N. Tait, T. F. de Lima, M. A. Nahmias, H. B. Miller, H. T. Peng, B. J. Shastri, and P. R. Prucnal, "Silicon photonic modulator neuron," Phys. Rev. Appl. **11,** 064043 (2019).
17. J. Feldmann, N. Youngblood, C. D. Wright, H. Bhaskaran, and W. H. P. Pernice, "All-optical spiking neurosynaptic networks with self-learning capabilities," Nature **569,** 208 (2019).
18. T. Van Vaerenbergh, M. Fiers, P. Mechet, T. Spuesens, R. Kumar, G. Morthier, B. Schrauwen, J. Dambre, and P. Bienstman, "Cascadable excitability in microrings," Opt. Express **20,** 20292 (2012).
19. G. Van der Sande, D. Brunner, and M. C. Soriano, "Advances in photonic reservoir computing," Nanophotonics **6,** 561 (2017).
20. T. Inagaki, K. Inaba, T. Leleu, T. Honjo, T. Ikuta, K. Enbutsu, T. Umeki, R. Kasahara, K. Aihara, and H. Takesue, "Collective and synchronous dynamics of photonic spiking neurons," Nat. Commun. **12,** 2325 (2021).
21. E. Khoram, A. Chen, D. Liu, L. Ying, Q. Wang, M. Yuan, and Z. Yu, "Nanophotonic media for artificial neural inference," Photonics Res. **7,** 823 (2019).
22. L. R. Cortes, D. Onori, H. G. de Chatellus, M. Burla, and J. Azana, "Towards on-chip photonic-assisted radio-frequency spectral measurement and monitoring," Optica **7,** 434 (2020).
23. A. E. Willner, A. Fallahpour, K. Zou, F. Alishahi, and H. Zhou, "Optical signal processing aided by optical frequency combs," IEEE J. Sel. Top. Quant. **27,** 7700916 (2021).
24. I. Coddington, N. Newbury, and W. Swann, "Dual-comb spectroscopy," Optica **3,** 414 (2016).
25. Y. Doumbia, T. Malica, D. Wolfersberger, K. Panajotov, and Marc Sciamanna, "Optical injection dynamics of frequency combs," Opt. Lett. **45,** 435 (2020).
26. Y. Doumbia, T. Malica, D. Wolfersberger, K. Panajotov, and Marc Sciamanna, "Nonlinear dynamics of a laser diode with an injection of an optical frequency comb," Opt. Express **28,** 30379 (2020).
27. H. Francis, X. Zhang, S. Chen, J. Yu, K. Che, M. Hopkinson, and C. Jin, "Optical frequency comb generation via cascaded intensity and phase photonic crystal modulators," IEEE J. Sel. Top. Quantum Electron. **27,** 2100209 (2021).
28. A. Murakami, K. Kawashima, and K. Atsuki, "Cavity resonance shift and bandwidth enhancement in semiconductor lasers with strong light injection," IEEE J. Quantum Elect. **39,** 1196 (2003).
29. A. C Arevian, V. Kapoor, and N. N Urban, "Activity-dependent gating of lateral inhibition in the mouse olfactory bulb," Nat. Neurosci. **11,** 80 (2008).
30. Y. Zhang, S. Xiang, X. Guo, A. Wen, and Y. Hao, "The winner-take-all mechanism for all-optical systems of pattern recognition and max-pooling operation," J. Lightwave Technol. **38,** 5071 (2020).
31. W. Zou, B. Ma, S. Xu, X. Zou, and X. Wang, "Towards an intelligent photonic system," Sci. China Inf. Sci. **63,** 160401 (2020).
32. Y. Duan, L. Chen, H. Zhou, X. Zhou, C. Zhang, and X. Zhang, "Ultrafast electrical spectrum analyzer based on all-optical Fourier transform and temporal magnification," Opt. Express **25,** 7520 (2017).
33. X. Xie, Y. Dai, Y. Ji, K. Xu, Y. Li, J. Wu, and J. Lin, "Broadband photonic radio-frequency channelization based on a 39-GHz optical frequency comb," IEEE Photon. Technol. Lett. **24,** 661 (2012).
34. C. Xiang, J. Liu, J. Guo, L. Chang, R. Wang, W. Weng, J. Peters, W. Xie, Z. Zhang, J. Riemensberger, J. Selvidge, T. J. Kippenberg, and J. E. Bowers, "Laser soliton microcombs heterogeneously integrated on silicon," Science **373**, 99 (2021).
35. J. Liu, G. Huang, R. Wang, J. He, A. S. Raja, T. Liu, N. J. Engelsen, and T. J. Kippenberg, "High-yield, wafer-scale fabrication of ultralow-loss, dispersion-engineered silicon nitride photonic circuits," Nat. Commun. **12**, 2236 (2021).